\documentclass[12pt]{article}

\usepackage{graphicx}
\usepackage{natbib}
\usepackage[margin=0.9in]{geometry}

\title{Single and binary stellar progenitors of long-duration gamma-ray bursts}

\author{{Dorottya Sz\'{e}csi}$^{1,2}$}
\date{$^1$~Astronomical Institute of the Czech Academy of Sciences, Fri\v{c}ova 298, 25165 Ond\v{r}ejov, Czech Republic\\
$^2$~School of Physics and Astronomy and Institute of Gravitational Wave Astronomy, University of Birmingham, Edgbaston, Birmingham B15 2TT, UK\\ \vspace{10pt}
E-mail: \textit{dszecsi@star.sr.bham.ac.uk, dorottya.szecsi@gmail.com}}

\begin{document}

\maketitle

\abstract{This review\footnote{XII Multifrequency Behaviour of High Energy Cosmic Sources Workshop\\
		12-17 June, 2017, Palermo, Italy} describes the most common theories behind long-duration gamma-ray burst progenitors. 
I discuss two astrophysical scenarios: the collapsar and the magnetar models. According to their requirements, the progenitor should be an envelope-free massive star with a fast rotating, collapsing iron core. Such an object, called a TWUIN star, may be produced by chemically homogeneous evolution either from a massive single star or a massive binary system. Various outcomes of this evolutionary path (e.g. supernova explosions and gravitational wave production) are also mentioned, and directions for future research are suggested. In the era of multi-messenger astronomy, my hope is to present a timely overview on how stellar astrophysicists are searching for progenitor models of long-duration gamma-ray bursts, and what they have found so far.
	}

\section{Introduction}

We are currently living in the era of multi-messenger astronomy. It is therefore especially important now to connect our theoretical understanding of gamma-ray burst (GRB) physics to the current and upcoming multi-wavelength and multi-messenger observations. 

This review is written from the perspective of a theorist. My goal is to present how stellar astrophysicists are searching for progenitor models of long-duration GRBs (LGRBs). These explosions are amongst the most energetic events ever detected---only the combined efforts of observers and theorists can bring us closer to solve the mysteries of their origins. 

At least two main categories of GRBs have been identified so far \citep{Blinnikov:1984,Paczynski:1986,1993ApJ...413L.101K,1996ApJ...471..915K,2005Natur.437..859H,2014ARA&A..52...43B,2016Ap&SS.361..155H}. These two categories are the short-duration, spectrally hard GRBs (SGRBs) and the long-duration, spectrally soft GRBs. They are, based on both their temporal and spectral properties, probably released by two different astrophysical sources \citep{2004IJMPA..19.2385Z}. SGRBs typically last for several tens of milliseconds. LGRBs typically last for dozens of seconds. Nonetheless, it is possible that there exist another category in between \citep[the so-called intermediate-duration GRBs,][]{2009Ap&SS.323...83H,2016Ap&SS.361..155H}, or that ultra-long GRBs should be categorized separately from the normally long ones \citep{2014ApJ...781...13L}.

From a theoretical point of view, the simplistic picture is that short-duration GRBs are produced by the merger of two compact objects. For a review on SGRBs written in the same vein as this one, see \citet{Szecsi:2017}.

As for LGRBs, it can be simply said that they are produced by massive stars at collapse. But the term \textit{massive star} is really broad \citep{Langer:2012}. It covers objects in the mass range of $\sim$9--300~M$_{\odot}$ (sometimes, hypothetically, even up to 10$^4$~M$_{\odot}$). They can be of solar composition (Z~$=$~Z$_{\odot}$), of low-metallicity (Z~$<$~Z$_{\odot}$) or completely metal-free (Z~$=$~0). Many of them rotate fast, and rotation may change the internal structure of a star significantly. Additionally, they may be born in a binary system with another star: in this case, the evolution is fundamentally influenced by the orbital separation and the mass ratio of the companions, as well as the actual mechanism with which they exchange mass. It is indeed a diverse set of objects that are summed up in the term \textit{massive star}, and surely not all of them end their lives as LGRBs.

This review is organized as follows. Sect.~\ref{sec:coll+magn} explains the special requirements that a massive stellar model should fulfill to be considered a LGRB progenitor. Sect.~\ref{sec:whatkind} describes those massive stellar models that fulfill said requirements. Sect.~\ref{sec:discussion} discusses the models, putting them into broader context of other explosions (e.g. supernovae, SGRBs, superluminous supernovae and gravitational wave emitting double black hole mergers). Sect.~\ref{sec:conclusion} concludes the review, pointing out some open questions in the field and possible directions for future research.

\section{Collapsars and Magnetars}\label{sec:coll+magn}

Massive stars is a broad term, and the task of the stellar physicists at hand is to chose those massive stars that are expected to explode as LGRBs. Indeed, to form a LGRB, a massive star has to fulfill some well-defined criteria. These criteria are usually coming from one of the two most commonly used astrophysical scenarios of LGRB explosions: the collapsar scenario and the magnetar scenario.

\subsection{The collapsar scenario}

Massive stars process hydrogen into helium, then helium into carbon and oxygen, then carbon and oxygen into other elements, and so on up until silicon. Silicon burns into iron, at which point nuclear fusion cannot maintain the star's hydrostatic stability anymore: iron cannot be burned into anything anymore via nuclear fusion because it is such a stable element that fusing it would \textit{require} energy instead of releasing energy.

The star's iron core collapses and falls in due to gravity. One may recall that this is how the usual textbook-explanation of a (core-collapse) supernova explosion continues \citep{Fryer:2004}: the iron core gets denser and denser, and eventually a proto neutron star (NS) forms in the middle. The material that is still falling in suddenly bounces back from the surface of the newly formed proto-NS. The outward bouncing gives rise to a shock-wave which may reach the surface and produce an emission of photons. This emission is what we may observe as a supernova lightcurve. If the iron core was more massive than $\sim$20~M$_{\odot}$, its self-gravity will very soon overcome the proto-NS's internal pressure, creating a compact object with such a strong gravitational field that nothing, not even particles and electromagnetic radiation, can escape from it. A black hole is formed. 

Let us now suppose that the bounce back is not very effective, and that only a weak shock-wave develops that is not able to reach the surface. Let us also suppose that the stellar core was rotating fast when the collapse started. In this case, we do not see a supernova lightcurve---the supernova is said to \textit{fail} and the outer layers fall into the black hole. However, equatorial outer layers of the fast-rotating core may retain a significant amount of angular momentum for an accretion disc to form around the central black hole. This set-up, called a collapsar, is thought to lead to a LGRB explosion because together with the accretion-disc, two jets also form, facilitating the gamma-ray production that is powered by the central black hole \citep{Woosley:1993,MacFadyen:1999,Yoon:2005,Woosley:2006}.

Summarizing the collapsar scenario of LGRBs, it requires
\begin{itemize}
\item an iron core that collapses,
\item a weak (or failed) supernova $\rightarrow$ material falls in,
\item a black hole in the middle (central engine),
\item fast rotation leading to an accretion disc and two jets.
\end{itemize}
Some LGRBs are observed together with a supernova explosion. In the context of the collapsar scenario, this may happen if the shock-wave reaches the surface leading to a regular supernova lightcurve---but if the core rotates fast, the accretion disc may still develop giving rise to a LGRB event accompanying the supernova. 

\subsection{The magnetar scenario}

Another promising scenario for the origin of long-duration GRBs is the proto-magnetar model. As opposed to the collapsar scenario where the central object is a black hole, the magnetar model supposes a fast rotating, magnetized proto-NS as the powering engine of the jet \citep{Usov:1992,Metzger:2011}. For such a proto-NS to form, the supernova definitely needs to be \textit{successful} and a significant amount of stellar material has to be ejected during the explosion. Thus, the proto-NS won't be massive enough to form a black hole, and forms a usual NS instead. A fast rotating, magnetized NS is called a magnetar. 

The magnetar scenario of LGRBs therefore require
\begin{itemize}
\item an iron core that collapses,
\item a successful, strong supernova $\rightarrow$ material ejected,
\item a fast rotating NS (magnetar) in the middle (central engine),
\item fast rotation leading to the magnetar, the accretion disc and two jets.
\end{itemize}
Additionally, to get a LGRB explosion powered by a magnetar we need a very strong magnetic field of $\sim$1e15~G. If the magnetic field is lower than that, \citet{Metzger:2015} suggested that we may see both a LGRB and a superluminous supernova together.

\subsection{Common features of the two scenarios}

While the collapsar scenario and the magnetar scenario suppose two different central engines, from the point of view of stellar progenitors, the two scenarios are actually not too different. The distinguishing feature is the strength of the supernova, but this is a complex problem that cannot be tackled by stellar evolution alone \citep{Fryer:2004,OConnor:2011,Ugliano:2012}. From the point of view of stellar evolution however, both scenarios require 
\begin{enumerate}
\item a massive star that evolves until an iron core forms, 
\item that the core rotates fast in the moment of the collapse, and 
\item that the progenitor star do not have a large (i.e. extended in physical space) envelope. 
\end{enumerate}
The last point is important because the jets should be able to penetrate through and break out of whatever envelope the progenitor star has, otherwise we would not see any electromagnetic (gamma) emission coming from the jet. In the following, we investigate what kind of massive stars would end their lives this way.

\section{What kind of star would die this way?}\label{sec:whatkind}

To fulfill either the collapsar or the magnetar scenario of LGRB formation, the stellar progenitor in question needs to form a fast rotating iron core while having only a very minor envelope. Red supergiants clearly do not qualify. Even if their core would somehow still rotate fast at collapse (e.g. due to the lack of magnetic coupling between core and growing envelope during the evolution), their envelopes are too extended for any jet to break through. Blue supergiants have been considered the progenitors of ultra-long GRBs \citep{Nakauchi:2013}. However, for the majority of regularly long GRBs, we need a stellar progenitor that is even more compact than a blue supergiant.

\subsection{Wolf--Rayet stars?}\label{sec:WR}

Classical Wolf--Rayet (WR) stars may come to mind. WR stars are hot ($\sim$100~kK) and relatively small ($\sim$10~R$_{\odot}$) objects with typical mass of around 20-30~M$_{\odot}$. Their stellar winds are optically thick, therefore we observe several broad emission lines in their spectra. Their surface is composed mainly of helium (as opposed to other stars having hydrogen-rich outer layers). So it seems that even if WR stars used to have a hydrogen-rich envelope when they were born, it has been somehow removed.

Textbooks indeed describe WR stars as the late stages of stellar evolution \citep{Kippenhahn:1990}. Stars originally more massive than 25~M$_{\odot}$ start their lives as O-type stars, then become blue supergiants, then red supergiant, then WR stars. Stars originally more massive than 40~M$_{\odot}$ also start out as O-type stars, then evolve to be blue supergiants, then luminous blue variables, then WR stars. This way of forming a WR star, however, supposes that the star was, for some period of time, either a red supergiant or a luminous blue variable, and then lost its envelope via its strong stellar wind.

Therefore, single WR stars are not predicted to rotate fast. Even if they were born rotating fast, they would have lost all their angular momentum with the removal of their envelope \citep{Brott:2011a,Groh:2014,Koehler:2015}. Unless a binary companion spins them up near the end of their lives \citep[see e.g.][and Sect.~\ref{sec:binaries}]{Fryer:2005}, they would not fulfill the criterion of a fast rotating core that is needed for both the collapsar and the magnetar scenarios.

\subsection{TWUIN stars}

Still discussing single stars, there is another evolutionary path to form a single massive star without a large envelope. Turns out, it even allows it to rotate fast. This evolutionary path, suggested first by \citet{Yoon:2005} and \citet{Woosley:2006} in the context of LGRB progenitors, is called chemically homogeneous evolution. 

The theory of stellar evolution predicts that if a low-metallicity ($<$~0.4~Z$_{\odot}$, but preferably even $<$~0.2~Z$_{\odot}$) massive star starts its life with fast rotation ($>$~300~km~s$^{-1}$), the internal mixing may be effective enough to keep the whole internal structure chemically homogeneous \citep{Maeder:1987,Yoon:2006,Meynet:2007,Brott:2011a,Koehler:2015,Szecsi:2015}. This means that the star will not develop a distinct core-envelope structure as 'normal' stars do, but stay compact, hot and homogeneous during all its life. Homogeneity implies that the surface composition reflects that of the core at every single time point of the evolution. That is, if the core is composed of helium for example (after hydrogen is exhausted from it), the surface is also mainly composed of helium. 

\begin{figure}
	\centering
	\includegraphics[angle=270,width=0.8\linewidth]{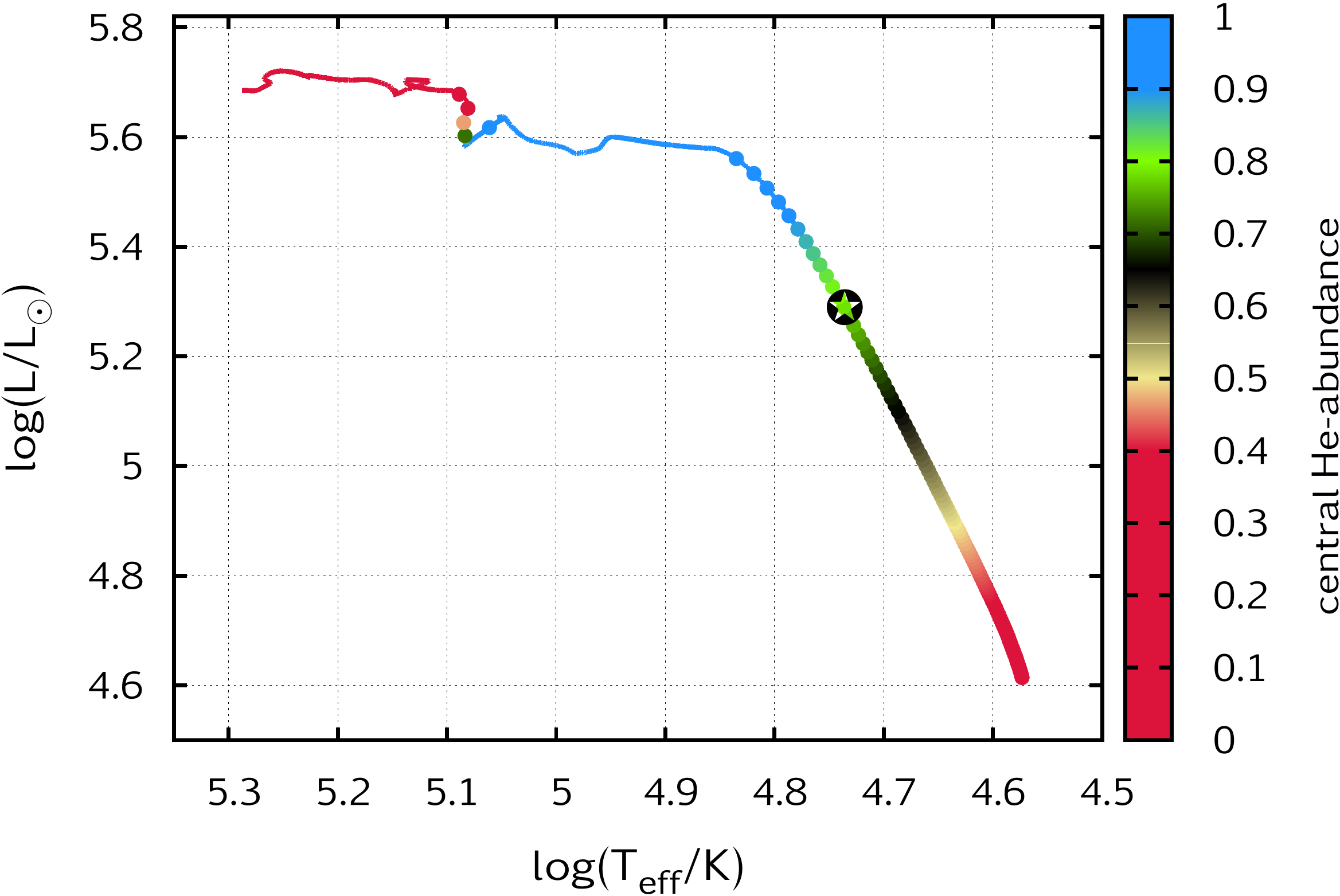}
	\caption{
		Hertzsprung--Russell diagram of a chemically homogeneous evolutionary model (colored straight line) and a TWUIN star (black five-point star symbol). The evolutionary model is taken from \citet{Szecsi:2015}. It has initial mass of M$_{\mathrm{ini}}$~$=$~20~M$_{\odot}$, initial rotational velocity of 450~km~s$^{-1}$ and metallicity of 0.02~Z$_{\odot}$. Its evolution starts in the right-bottom corner (zero-age main sequence) and proceeds toward higher luminosities and higher effective temperatures. Dots mark every 10$^5$~years of evolution: the clustering of dots at log~T$_{\mathrm{eff}}$~$\sim$~5.08 implies the core-helium burning phase (cf. color shading). The computations were followed until carbon is exhausted from the core. 
		An object predicted to lie on a chemically homogeneous stellar track that has optically thin wind, is called a TWUIN star. For example, the TWUIN star marked in this diagram has the following properties: M~$=$~19.8~M$_{\odot}$, R~$=$~4.95~R$_{\odot}$, 
		surface helium mass fraction Y$_{\mathrm{S}}$~$=$~0.75, core helium mass fraction Y$_{\mathrm{C}}$~$=$~0.78,
		mass loss rate $\dot{\mathrm{M}}$~$=$~$-$6.89 [log~M$_{\odot}$~yr$^{-1}$], equatorial rotational velocity at the surface v$_{\mathrm{rot}}$~$=$~650~km~s$^{-1}$ and escape velocity v$_{\mathrm{esc}}$~$=$~1237~km~s$^{-1}$. The optical depth in the wind \citep[estimated following][]{Langer:1991} is $\tau$~$=$~0.2.
		}\label{fig:hrd}
\end{figure}

Notice the criterion of fast rotation because here is where the role of low-metallicity comes into the picture: with high-metallicity comes a strong stellar wind that removes angular momentum, putting an end to the chemically homogeneous evolution. But if the metallicity is low, the wind is less strong and the star may keep most of its initial angular momentum. Thus, it can stay chemically homogeneous during all its life, thus \textit{avoiding} the supergiant phase and the luminous blue variable phase. Fig.~\ref{fig:hrd} shows such a stellar evolutionary model.

The term 'chemically-homogeneous evolution' means a type of stellar evolution: that is, stellar models with tracks covering the hot side (T$_{\mathrm{eff}}$~$\sim$~35-120~kK) of the Hertzsprung--Russell diagram. On the other hand, as \citet{Szecsi:2015} has shown, an  \textit{object} predicted to lie on one of the chemically homogeneous stellar tracks\footnote{~
	To be very precise, not all the objects that lie on chemically homogeneous evolutionary tracks are stars with transparent wind. In some particular cases, the object may have optically thick wind. Examples include the very massive (>~100~M$_{\odot}$) homogeneous models of \citet{Brott:2011} and \citet{Koehler:2015} with Z~$\sim$~0.4~Z$_{\odot}$ which are WR~stars. These models are not predicted to be progenitors of LGRBs though.
	
	So we can simply say that TWUIN stars are chemically homogeneously evolving models, but not the other way around: technically, chemically homogeneous evolution does not always produce TWUIN stars. When we talk about LGRB progenitors, however, we can safely omit this distinction because only those chemically homogeneously evolving models rotate fast enough at core collapse that have been TWUIN stars during their lives. 
	} may be called a TWUIN star.

\subsection{What do we know about TWUIN stars?}\label{sec:whatdoweknow}

The abbreviation TWUIN stands for Transparent Wind Ultraviolet INtense. The distinguishing properties of these stars are in the name: they have weak, optically thin stellar winds, and they are hot thus emitting radiation in the UV band. What is not in the name but important: they are massive stars with low metallicity (Z~$<$~0.2~Z$_{\odot}$), and they rotate fast. 

TWUIN stars are not WR stars \citep{Szecsi:2015b}. They are, for any means, massive stars with surface composition and temperature typical for WR stars, but with wind optical depth typical for O-type stars. 

Until now, TWUIN stars are mainly theoretical predictions. Nonetheless, they offer us huge scientific potential in terms of explaining LGRB progenitors. Several authors provided stellar models that evolve chemically homogeneously, showing that they indeed fulfill the criteria of the collapsar scenario \citep{Yoon:2005,Woosley:2006,Yoon:2006,Cantiello:2007,Yoon:2012,Szecsi:2016}. Most of these authors did not consider the magnetar scenario though, the details of which were significantly elaborated on only later by \citet[][]{Metzger:2011}. 

A distinguishing feature of TWUIN stars is that, according to the simulations, they produce a huge amount of ionizing photons. It has been therefore speculated by \citet{Szecsi:2015b} that we might have indeed gathered indirect evidence for their existence. As reported by \citet{Kehrig:2015}, low-metallicity (Z~$\sim$~0.02~Z$_{\odot}$) dwarf galaxy I~Zwicky~18 is currently undergoing star formation. They observed very high levels of photoionzation in this galaxy which could not be explained either by WR stars, nor by X-ray binaries and shocks from supernova remnants (the most common sources of ionizting photons). While they suggest that the presence of very massive metal-free stars could possibly explain the observations, \citet{Szecsi:2015} argues that there is no need for supposing metal-free stars being present in a metal-poor galaxy. Stellar models computed with the exact composition of I~Zwicky~18 predict several TWUIN stars. They estimated the ionization produced by a population of TWUIN stars, and found that it can explain the observations of \citet{Kehrig:2015} quite well. 

This means two things. First, if TWUIN stars are responsible for the ionization observed in I~Zwicky~18 \citep[or in any other low-metallicity star-forming galaxy, cf.][]{Shirazi:2012}, \textit{we should expect LGRBs exploding there at some point} \citep[cf.][]{Szecsi:2015b}. 

Second, we should focus our attention to the look of TWUIN stars. This should be done both theoretically (simulating their atmospheres and spectra for example) and observationally (conducting searches that look for them in neighboring low-metallicity galaxies). Some pioneer studies in the topic include \citet{Tramper:2013,Szecsi:2015,Hainich:2017,Hamann:2017}, but there is clearly much more to be done. 

\subsection{Binary progenitors of LGRBs}\label{sec:binaries}

Most massive stars are in a binary system \citep{Sana:2012}. Therefore, while it is unquestionably useful to know how single stars evolve, we must pay attention to what happens if they interact with a companion. 

A fast-rotating massive star without an envelope may be formed by binary interaction. \citet{Fryer:2005} created such  a model via a common-envelope in-spiral. The final product that explodes in their model is a fast-rotating WR star. 

\citet{Cantiello:2007} suggested another way: stripping down the envelope at the early phases of evolution in a close binary system, making the stripped star spin up. Then, it evolves chemically homogeneously (as they applied low-metallicity, Z~$\sim$~0.2~Z$_{\odot}$ models). Technically, their chemically homogeneously evolving LGRB progenitor could be considered a TWUIN star.

Although focusing on progenitors of gravitational-wave emission, \citet{Marchant:2016} and \citet{Marchant:2017} created binary models in which both companions are expected to produce a LGRB (with or without an accompanying supernova) before turning into black holes that eventually merge. Again, as this evolutionary path also involves chemically homogeneous evolution and low-metallicity (Z~$\sim$~0.02~Z$_{\odot}$), these objects should be considered TWUIN binaries.

As for high-metallicity, observational results of \citet{Vergani:2015} and \citet{Japelj:2016} imply that some (but not all) LGRBs form with Z~$\sim$~Z$_{\odot}$. Since single stars at high-metallicity cannot retain enough angular momentum for producing a collapsar or magnetar due to their strong mass-loss (cf.~Sect.~\ref{sec:WR}), the only stellar evolutionary channel that may lead to such an explosion should necessarily involve massive binaries. More research on massive binary stars is needed to provide the missing connection between observations and theory of high-metallicity progenitors of long-duration GRBs.

\section{Discussion}\label{sec:discussion}

\subsection{Implications for cosmology}\label{sec:cosmology}

One can hand-wavingly say that TWUIN stars are Population~II massive stars. Indeed, we expect them to be present in the early Universe in large number
\citep{Yoon:2006,Szecsi:2015} as part of the second and subsequent generations of stars. More studies are needed to decide how (or if) they contributed to the re-ionization history of the Universe \citep[cf.][]{Meszaros:2010,Amati:2017}.

\subsection{Other explosions from single TWUIN stars}

Not all single TWUIN stars are predicted to explode as LGRBs. In fact, only those are considered progenitors of a collapsar or a magnetar that have core mass of 10~M$_{\odot}$~$<$~M$_{\mathrm{core}}$~$<$~40~M$_{\odot}$. They may or may not be accompanied by a supernova (of type I~b/c) depending on the nature of the core-collapse and the explosion \citep{Woosley:2006b,Yoon:2006}. On the other hand, it may happen that, in a case both a LGRB and a supernova occurs, we only see the latter because the jets are not aligned with the line-of-sight.

As for TWUIN stars with cores more massive than 40~M$_{\odot}$, we expect them to undergo pair-instability. Metal-poor massive stars are predicted to encounter an instability during neon or oxygen burning. It is due to pair-creation---hence the name. Its consequence is that the subsequent evolution may never occur. These stars collapse even \textit{before} an iron core forms \citep{Burbidge:1957,Langer:1991,Heger:2003,Langer:2007,Kozyreva:2014}. 

The outcome of the collapse may vary. In case of very massive cores (above $\sim$133~M$_{\odot}$), the star collapses directly into a black hole without an explosion. In case of less massive cores, the collapse could be stopped and reversed by the nuclear energy release of explosive oxygen burning, and a pair-instability supernova event would happen. This disrupts the whole star leaving no remnant. For more detailed discussions on the subject, we refer to \citet{Herzig:1990,Heger:2002,Langer:2007,Woosley:2007,Kasen:2011,Chatzopoulos:2012,Dessart:2013,Kozyreva:2014,Sukhbold:2016,Szecsi:2016}.

\subsection{Other explosions from binaries}\label{sec:otherbinaries}

Before discussing the results of existing studies on binary evolution, there is an important, but sometimes overlooked aspect of the methods these studies are carried out with.

Theoretical studies on binary evolution usually fall into one of two categories: either detailed evolutionary simulation, or population synthesis computations. The first means that the stellar structures of both companions are simulated individually, and their joint evolution is followed in detail, possibly even including their mass exchange phase. This is an unquestionably challenging task. Nonetheless, several works have explored massive binaries with this method \citep[e.g.][]{Cantiello:2007,Eldridge:2008,deMink:2009b,deMink:2009,Yoon:2010,Yoon:2015,Marchant:2016,Song:2016,Marchant:2017}, some of them motivated by the potential they have in explaining gravitational wave emissions. These works could, however, only cover some limited part of the vast parameter space in metallicity, orbital separation, mass ratio and mass transfer efficiency. 

On the other hand, studies involving binary population synthesis \citep[e.g.][]{Belczynski:2002,Dominik:2012,Dominik:2013,Belczynski:2016,deMink:2016,Mandel:2016} are, while essentially important to predict the outcome from large populations of massive stars, not computing detailed models. They \textit{rely on} detailed models, of which only a limited set has ever been created (and implemented into said population synthesis studies). Therefore, an essentially important future direction of research is to compute detailed binary models covering larger and larger parts of the parameter space, and to implement their results into population synthesis codes.

Below I shortly recite the results of detailed simulations, especially those that involve TWUIN stars. As mentioned in Sect.~\ref{sec:binaries}, several binary evolutionary models operate with chemically homogeneous evolution. \citet{deMink:2009b} showed that two low-metallicity massive stars in a very close orbit ($<$~3~days) can produce a double black hole system that merge eventually and produce gravitational wave emission. \citet{Marchant:2016} computed a large set of such models, finding that this scenario of gravitational wave production prefers low-metallicity. See \citet{Perna:2016}, \citet{Bagoly:2016} and \citet{Szecsi:2017} for discussions on this scenario being accompanied by a SGRB.

\citet{Marchant:2017} extended the scenario to involve the merger of a neutron star with a black hole. They expect that these systems produce the most massive accreting stellar black holes which are the brightest sources of ultra-luminous X-ray emission. If the merger of a neutron star and a black hole happens within the lifetime of the Universe, we may observe it as SGRBs \citep{Blinnikov:1984,Paczynski:1986,Berger:2014,Paschalidis:2015,Szecsi:2017}.

\section{Summary and outlook}\label{sec:conclusion}

Theories of single and binary stellar evolution that predict long-duration gamma-ray burst progenitors have been reviewed. The astrophysical scenario called a collapsar involves a fast rotating, massive stellar core that is collapsing and thus forms an accretion disc. The central black hole powers the jets. It may be accompanied by a supernova of type I~b/c. The magnetar scenario is similar, but in it the central object is a fast rotating, magnetized neutron star. Both scenarios require a star without an envelope so the jets can break out.

Single stellar models that predict fast rotating, collapsing cores and no envelope may form via chemically homogeneous evolution. This evolutionary scenario only happens at low-metallicity, when the model star rotates fast. Fast rotation may be the result of the star formation process, or the interaction with a binary companion. In the latter case, both of the companions may explode as LGRBs, albeit with a time difference of some ten to hundred thousands of years.

An object corresponding to a given time point of a chemically homogeneous evolutionary track is called a TWUIN star. TWUIN stars have transparent wind, and produce intense radiation in the UV. It has been speculated that they may be responsible for the high ionizing flux in dwarf galaxies.

Not all TWUIN stars explode as LGRB. Some predict pair-instability supernovae, others supernovae of type~I~b/c. Chemically homogeneous evolution has been used to model the stellar progenitor of double black hole mergers leading to gravitational wave emission. 

Several important directions for future research have been pointed out during my review, namely:
\begin{enumerate}
	\item to simulate the atmospheric structure and spectra of TWUIN stars (Sect.~\ref{sec:whatdoweknow}); 
	\item to observe the massive star population of low-metallicity, star-forming dwarf galaxies in the local Universe to test the models of stellar evolution and construct spectral catalogue of low-metallicity massive stars (Sect.~\ref{sec:whatdoweknow}); 
	\item to expand the parameter space that detailed binary evolutionary simulations cover (Sects.~\ref{sec:binaries} and \ref{sec:otherbinaries}); 
	\item to implement the outcome of these new evolutionary simulations into population synthesis studies (Sect.~\ref{sec:otherbinaries});
	\item to tackle the open question how TWUIN stars (i.e. Population~II massive stars) contributed to the re-ionization of the early Universe (Sect.~\ref{sec:cosmology}).
\end{enumerate}

\section*{Acknowledgements}

D.Sz.\ was supported by the Czech Grant nr.\ 16-01116S GA \v{C}R.
\bibliographystyle{aa} 
\bibliography{References} 

\bigskip
\bigskip
\noindent {\bf DISCUSSION}

\bigskip
\noindent {\bf Dmitry Bisikalo:} What is the source of fast rotation of TWUIN stars?

\bigskip
\noindent {\bf Dorottya Sz\'ecsi:} A massive star may start its life with fast rotation either due to the birth cloud spinning fast, or some early interaction with a binary companion \citep[for example,][]{Cantiello:2007}. If the star in question has low metallicity, the internal mixing may be effective enough to keep the whole internal structure chemically homogeneous, thus creating a TWUIN star.



\end{document}